\documentstyle[epsf,amstex,epsfig]{mn}
\loadboldmathitalic 
\title []{Citation measures and impact within astronomy}

\author[Pearce, F.R.]
{Frazer R. Pearce$^{1,3}$\\
$^{1}$Department of Physics \& Astronomy, University of Nottingham,
NG7 2LU \\
$^{3}$email:Frazer.Pearce@@nottingham.ac.uk\\
}
\date{\today}
\def\lesssim{\mathrel{\hbox{\rlap{\hbox{\lower4pt\hbox{$\sim$}}}\hbox{$<$}}}}
\def\gtrsim{\mathrel{\hbox{\rlap{\hbox{\lower4pt\hbox{$\sim$}}}\hbox{$>$}}}}
\def\ion#1#2{#1$\;${\small\rm\@roman{#2}}\relax}
\def\etal{{\it et al.\thinspace}}

\begin{document} 
\maketitle 
\begin{abstract}
By utilising the inbuilt citation counts from NASA's astrophysics
data system (ADS) I derive how many citations refereed articles
receive as a function of time since publication. 
After five years, one paper in a hundred has accumulated 91
or more citations, a figure which rises to 145 citations after ten
years. By adding up the number of citations active researchers have
received over the past five years I have estimated their relative
impact upon the field both for raw citations and citations weighted by
the number of authors per paper.
\end{abstract}

\section{Introduction}\large

What makes a good paper? No objective measure is ever going to be
perfect but being cited in another paper at least indicates that the
work has been noticed and thought to be worth mentioning. Papers with
many citations are, in general, likely to be more useful and
interesting than those that sink without a trace. This is much the
same system as that used successfully by fast internet search engines
to score sites so that they can provide a list ordered by usefulness;
those sites which many people link to get a high score and appear near
the top of the returned list. Such a system is admittedly far from
perfect, as it can be influenced by many factors such as having a
large number of friends who cite you, self-citing your own papers
excessively or producing a very good paper that concludes an avenue of
research in such a way that it doesn't form the basis for a large body
of subsequent endeavour.  That said, there is undoubtedly a trend;
papers with high citation counts tend to be better written, more
interesting and useful than those that never get referred to again.

In section 2 I first look at citation classics, the 1000 most cited
astronomy papers according to the ADS. Next I examine the citation
counts for refereed astronomy papers published since 1970. From this
the number of citations received by 1-in-10, 1-in-100 and 1-in-1000
papers can be obtained, and are shown in table ~1. The number of
citations received by papers of a specified age is also shown, 
where the age stated defines the centre of a one year range
(so 2 years ago means papers published between 18 and 30 months ago).

When can a researcher be said to have a high ongoing impact on the field?  
This question is even harder to answer, but a sequence of well cited
papers would be a good start. In section 3 I suggest two
measures of current impact on the field which are calculated by
tallying up a researchers total citations or normalised total
citations over a rolling five year interval. I calculate these numbers
for a random sample of over 5000 astronomers with current publications
in order to produce likelihood curves.

\section{Citation classics}\large

Quite how a paper passes into folklore and becomes a {\it citation
classic} is difficult to determine. To enter the ranks of the 1000 most
cited astronomical papers in the ADS archive 
requires a paper to have obtained 257 citations (as of november 2003, when
the archive contained 439746 papers). For the purposes of this study,
papers are defined to be those found in the astronomy/planetary ADS
archive, published in all refereed journals. Citations counts are
those returned by ADS. It should be
noted that these citation counts are not complete, with some
references omitted because the citing journal is not within the ADS
database or because an older article has only been scanned at present.
Inaccuracies in the citing papers reference list can also lead to
missed citations. The ADS does contain complete reference lists for
all the major astrophysics journals back to issue 1.

Figure~1 shows the distribution by publication year of the 1000 most
cited astronomical papers. This peaks around 1985, a year which
contributes nearly 50 papers to the total. It takes around a decade
for the number of citation classics per year to rise to over 30, a
figure which tallies well with the 5 year timescale required to reach
a maximum citation rate followed by a slow decline (Abt 1981).

The oldest of these (Chandrasekhar
1943), `Stochastic Problems in Physics and Astronomy', 
is a true classic, with over 900 citations, whilst the second
oldest (Bondi \& Hoyle 1944), `On the mechanism of accretion by stars'
will most likely soon drop out of the list as it `only' has 284
citations. Papers published this long ago are likely to have many
contemporary citations missed as they have not yet been entered into
the archive. The most recently published additions to the list are Ahmad
\etal (2001) -- (for neutrino measurements) and Freedman \etal (2001) --
(measuring the Hubble constant), with just over 300 citations each.

Many of the most highly cited papers in this list are measurements of
fundamental parameters - Kurucz 1979, `Model atmospheres for G, F, A,
B, and O stars', Anders \& Grevesse 1989, `Abundances of the elements
- Meteoritic and solar', Landolt 1992, `UBVRI photometric standard
stars in the magnitude range 11.5-16.0 around the celestial equator',
Savage \& Mathis, 1979 `Observed properties of interstellar dust' and
Draine \& Lee, 1984, `Optical properties of interstellar graphite and
silicate grains', are all in the top 10 most cited papers.

\begin{figure}
\begin{center}
\leavevmode \epsfysize=8cm \epsfbox{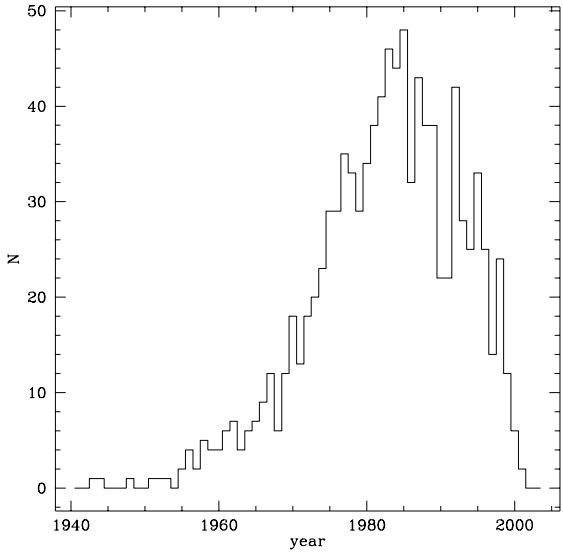}
\end{center}
\caption{Histogram showing the publication year of the 1000 most cited
astronomical papers. As of november 2003 an individual paper required
257 citations to make it onto this plot
}
\end{figure}

\begin{table}
\begin{tabular}{lcccc}
Likelihood		&0.5	&0.1	&0.01	&0.001	\\
\hline
1 year ago		&0	&5	&16	&39	\\
2 years	ago		&1	&11	&35	&101	\\
3 years	ago		&2	&17	&59	&173	\\
4 years	ago		&4	&22	&74	&219	\\
5 years	ago		&4	&26	&91	&253	\\
10 years ago		&6	&41	&145	&344	\\
20 years ago		&6	&47	&188	&594	\\
1970+			&3	&35	&138	&396	\\
\hline
citations (2 papers)	&16	&231	&1063	&2463	\\
normalised citations (2)&3	&41	&168	&357	\\	
citations (5 papers)	&61	&382	&1551	&2597	\\
normalised citations (5)&12	&74	&229	&371	\\	
\hline
\end{tabular}
\caption{The number of citations required to cross successive
power-of-ten likelihoods for papers published a specified number of
years (plus six months) ago. 
The last four rows show the required number of citations and
normalised citations (see text) required {\it per researcher} to reach
the same likelihoods.
\label{tab1}}
\end{table}

How many citations do more normal papers receive? Using the counts in
the ADS archive I have calculated the likelihood of a paper achieving
a specified number of citations for all papers in the archive
published since 1970. I have also calculated these likelihoods for
papers of a specified age, plus or minus six months. Nowadays there
are roughly 15,000 papers published per year, falling to 10,000 and
8,500 ten and twenty years ago. Figure~2 shows these likelihoods for a
range of ages as well as the long term average (bold line) and the
required number of citations to cross certain thresholds are given in
table~1. One-in-ten papers published five years ago has now received
26 citations, with one-in-100 getting more than 91 and one-in-1000
more than 253 citations.

\begin{figure}
\begin{center}
\leavevmode \epsfysize=8cm \epsfbox{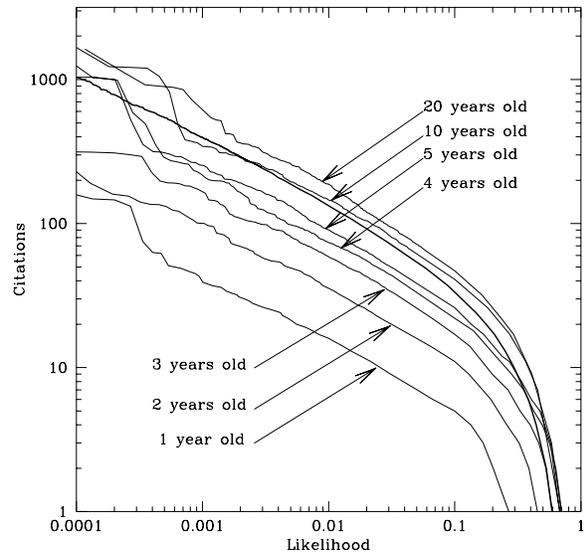}
\end{center}
\caption{Likelihood of a paper obtaining a specified number of
citations for all papers published since 1970 (bold line) and those
published a specified number of years ago. The numbers of citations
required to breach each likelihood decade are given in table 1.
}
\end{figure}

\section{Good researchers}\large

Defining a rating system for researchers is a subject fraught with
danger as any single proposed scheme is bound to contain
inconsistencies. Here I attempt to come up with a scheme that provides
a general guideline. An individual high score on either of the metrics
should be subject to further examination and should only be treated as
an indication - a researcher in the top few percent of both metrics is
likely to be far more widely known than one in the bottom half.

The basis of both schemes described here is to only consider papers
published in refereed journals within a rolling five year time
interval. This timespan is set to start five and a half years before
the current time and finish six months ago (as there are almost no
citations in the first six months after publication). The citation
score is then calculated by summing up the total number of citations
for those papers up to the current date using the automatic facility
built in to ADS. This procedure is achieved by going to the page;
{\rm http://ukads.nottingham.ac.uk/abstract\_service.html}, typing 
a name in the author field and (in November 2003) the dates 06/1998
and 05/2003 as the range of publication dates. The ``select references
from'' field needs to be changed to ``All refereed journals'' and
sorting done by citation count. This should return figures for the
number of papers found and the total number of citations for those
papers. There will also be an ordered list of the papers, each with
its own citation count, although this information is not used here.
A second measure, the normalised citation score is calculated by
weighting each paper by the number of authors, reducing the
impact of large collaborations which often produce a large number of
papers. No attempt is made to remove self citations, however
gratuitous (Pearce \etal 2000).

The difficult part of this study is obtaining a useful list of names
to feed to the algorithm. I have taken all the unique names plus first
initial contributing two or more items to the ADS for authors
whose surname starts with A, B or C. In addition,  I have
also disabled synonym matching on author names,
which avoids pattern matching on middle initials and phonetic
pronunciation matching. This procedure will combine authors with
names such as Martin, A. S. and Martin, A. C. together, but such very
similar names appear to be rare, at least when both are successful. I
have checked that results near the top of the study suffer no
detectable contamination. 

The final list of 18,346 names is automatically passed to the ADS
server which returns citation and normalised citation figures as
detailed above. Not all these astronomers have published during the five
year period, as the source list of names spans the entire database. In
total 5136, or roughly 30\% of the authors have published two or more
refereed papers 
in the study period. Figures~3 \& 4 below show the likelihood of
achieving different citation and normalised citation counts
respectively, both for all the authors with two or more citations and
for the 2467 active researchers who published at least one paper a year
over the study period.

\begin{figure}
\begin{center}
\leavevmode \epsfysize=8cm \epsfbox{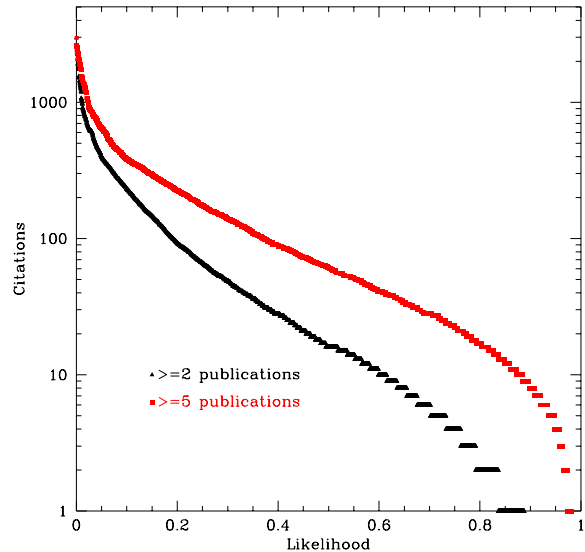}
\end{center}
\caption{Likelihood of an author achieving more than a specified
number of citations within a recent 5 year window. The lower curve is
for all authors publishing 2 or more papers in the
interval, the upper curve is for {\it active} researchers with 5 or
more recent papers.
}
\end{figure}

\begin{figure}
\begin{center}
\leavevmode \epsfysize=8cm \epsfbox{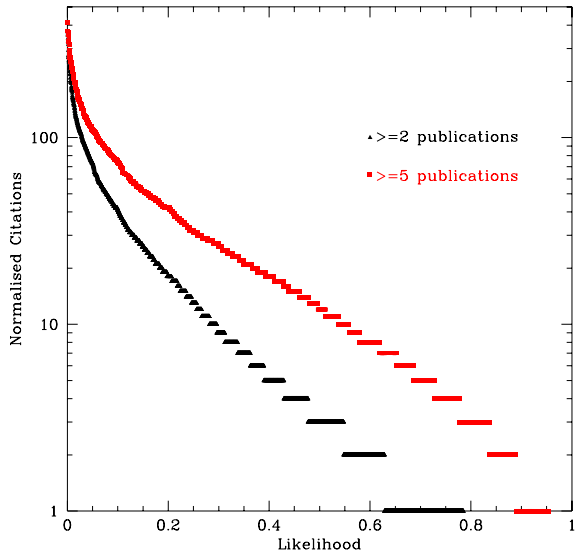}
\end{center}
\caption{Likelihood of an author achieving more than a specified
number of normalised citations within a recent 5 year window. The lower curve is
for all authors publishing 2 or more papers in the
interval, the upper curve is for {\it active} researchers with 5 or
more recent papers.
}
\end{figure}

As table~1 lists, ten percent of astronomers with two or more refereed
papers in the last five years received more than 231 citations, with
this figure rising to 382 for active researchers. Similarly,
one-in-ten publishing astronomers receive more than 40 normalised
citations in the same period. 

\section{Conclusions}\large

By extracting citation counts from the ADS I have produced a list of
the 1000 most cited astronomical papers and produced likelihoods for
papers receiving a certain number of citations as a function of the
number of years that have elapsed since publication. 

In this short work I have calculated two easily determined measures of
success within the astronomical community. Although these numbers
should be treated carefully they do at least provide some indication
of the impact of a particular researcher on the field and relative to
their contemporaries. The numbers derived here are simply reproducible
for any given person and comparisons can be made against the averages
for the community as a whole. 

\section*{Acknowledgements}
This research has made use of NASA's Astrophysics Data System
and arose out of discussion with Duncan Forbes over relevant
measures of success for use in funding battles.

\section*{References}
\noindent Abt, H.~A.\ 1981, PASP, 93, 207 

\noindent Ahmad, Q.~R.~et al.\ 2001, Physical Review Letters, 87, 71301 

\noindent Anders, E.~\& Grevesse, N.\ 1989, GCA, 53, 197 

\noindent Bondi, H.~\& Hoyle, F.\ 1944, MNRAS, 104, 273 

\noindent Chandrasekhar, S.\ 1943,  Reviews of Modern Physics, 15, 1 

\noindent Draine, B.~T.~\& Lee, H.~M.\ 1984, ApJ, 285, 89 

\noindent Freedman, W.~L.~et al.\ 2001, ApJ, 553, 47 

\noindent Kurucz, R.~L.\ 1979, ApJS, 40, 1 

\noindent Landolt, A.~U.\ 1992, AJ, 104, 340 

\noindent Pearce, F.~R., Thomas, P.~A., Couchman, H.~M.~P., \& Edge, A.~C.\ 2000, 
MNRAS, 317, 1029 

\noindent Savage, B.~D.~\& Mathis, J.~S.\ 1979, ARAA, 17, 73 

\end{document}